\documentclass[submission,copyright,creativecommons]{eptcs}
\providecommand{\event}{ACT 2021}
\usepackage{underscore}           

\usepackage{subfiles}
\usepackage[utf8]{inputenc}
\usepackage{fancyhdr}
\usepackage{amsmath}
\usepackage{amssymb}
\usepackage{amsthm}
\usepackage[utf8]{inputenc}
\usepackage{amsthm}

\usepackage[mathscr]{euscript}
\usepackage{ebproof}
\usepackage{upgreek}
\usepackage{nicematrix}
\usepackage{stmaryrd}
\usepackage{quiver}

\theoremstyle{definition}
\newtheorem{theorem}{Theorem}
\newtheorem{definition}[theorem]{Definition}

\newtheorem{example}[theorem]{Example}

\newtheorem{proposition}[theorem]{Proposition}

\DeclareMathSymbol{'}{\mathord}{operators}{"3A}

\hypersetup{colorlinks=true}

\newcolumntype{L}{>$l<$}
\newcolumntype{R}{>$r<$}
\newcolumntype{C}{>$c<$}

\newcommand{\define}[1]{{\bf \boldmath{#1}}}

\newcommand{\msf}[1]{\mathsf{#1}}
\newcommand{\mbb}[1]{\mathbb{#1}}
\newcommand{\msc}[1]{\mathscr{#1}}
\newcommand{\mrm}[1]{\mathrm{#1}}
\newcommand{\mtt}[1]{\mathtt{#1}}

\newcommand{\rtail}{\rightarrowtail}
\newcommand{\xr}[2]{\xrightarrow[#2]{#1}}
\newcommand{\ra}{\Rightarrow}
\newcommand{\op}{\msf{op}}

\newcommand{\ess}{\msf{S}}
\newcommand{\T}{\msf{T}}
\newcommand{\Th}{\msf{Th}}
\newcommand{\Set}{\mrm{Set}}
\newcommand{\Pos}{\mrm{Pos}}

\newcommand{\Sub}{\mrm{Sub}}
\newcommand{\CHA}{\mrm{CHA}}
\newcommand{\Topos}{\mrm{Topos}}

\newcommand{\Prop}{\msf{Prop}}
\newcommand{\Type}{\msf{Type}}

\newcommand{\Thy}{\mrm{Thy}}

\newcommand{\F}{\msf{F}}

\newcommand{\FF}{\mrm{F}}
\newcommand{\I}{\mrm{I}}
\newcommand{\EE}{\mrm{E}}

\newcommand{\pow}{\msc{P}}
\newcommand{\sfy}{\msf{y}}
\newcommand{\wand}{\mbox{--\!$\ast$}}
\newcommand{\interp}[1]{\llbracket #1 \rrbracket}

\newcommand{\E}{\mtt{E}}
\newcommand{\V}{\mtt{V}}
\newcommand{\Hom}{\mtt{Hom}}

\newcommand{\tte}{\mtt{e}}

\newcommand{\ttf}{\mtt{f}}
\newcommand{\ttg}{\mtt{g}}

\newcommand{\ttC}{\mtt{C}}
\newcommand{\ttr}{\mtt{R}}
\newcommand{\tts}{\mtt{S}}
\newcommand{\ttt}{\mtt{T}}

\newcommand{\N}{\mtt{N}}
\newcommand{\PP}{\mtt{P}}
\newcommand{\tto}{\mtt{out}}
\newcommand{\tti}{\mtt{in}}
\newcommand{\ttz}{\mtt{0}}

\newcommand{\q}{\mtt{q}}

\newcommand{\var}{\mtt{var}}
\newcommand{\app}{\mtt{app}}
\newcommand{\lam}{\mtt{lam}}
\newcommand{\deff}{\mtt{def}}

\title{Native Type Theory}
\author{Christian Williams
\institute{University of California, Riverside, US}
\email{cbwill.math@gmail.com}
\and
Michael Stay
\institute{Pyrofex Corporation, Utah, US}
\email{stay@pyrofex.net}
}
\def\titlerunning{Native Type Theory}
\def\authorrunning{C. Williams \& M. Stay}

\begin{document}
\maketitle

\begin{abstract}
    Native type systems are those in which type constructors are derived from term constructors, as well as the constructors of predicate logic and intuitionistic type theory. We present a method to construct native type systems for a broad class of languages, $\lambda$-theories with equality, by embedding such a theory into the internal language of its topos of presheaves. Native types provide total specification of the structure of terms; and by internalizing transition systems, native type systems serve to reason about structure and behavior simultaneously. The construction is functorial, thereby providing a shared framework of higher-order reasoning for many languages, including programming languages.
\end{abstract}








\section{Introduction}
\label{sec:intro}
Type theory is growing as a guiding philosophy in the design of programming languages. However in practice, type systems are heterogeneous, and there are no standard ways to reason across languages. We present a functorial method to enhance a language with ``its own internal logic'', using tools and ideas of category theory.

Categorical logic unifies languages: virtually any formalism, from monoids to dependent type theory, can be modelled by structured categories \cite{jacobs}. By doing so, we inherit a wealth of tools from category theory. In particular, we can generate expressive type systems by composing two known ideas.
\[\begin{tikzcd}
	{\lambda\mathtt{theory}} & {\mathtt{topos}} & {\mathtt{type \; system}}
	\arrow["{\mathscr{P}}", from=1-1, to=1-2]
	\arrow["{\mathscr{L}}", from=1-2, to=1-3]
\end{tikzcd}\]

The first is the \textit{presheaf construction} $\mathscr{P}$ \cite[Ch.\ 8]{awodey-CT}; it preserves product, equality, and function types. The second is the \textit{language of a topos} $\mathscr{L}$ \cite[Ch.\ 11]{jacobs}. The composite is 2-functorial, so that translations between languages induce translations between type systems.

Note --- This idea is quite simple; in fact it was considered more than fifty years ago by Scott \cite{scott}. Native type theory simply gives a name to the language of presheaves on theories; we aim to demonstrate its utility, and advocate for real-world application of categorical logic.

The type system is \textit{native} in the sense that type constructors are derived from term constructors, plus those of predicate logic and (co/inductive) intuitionistic type theory. For example, the following predicate on processes in a concurrent language (ex.\ \ref{ssec:rho}) is effectively a compile-time firewall.
\[\begin{array}{l}
    \msf{sole.in}(\alpha) \; := \;\;\; \nu\mtt{X}.\; (\tti(\alpha,\N\to \mtt{X})\;\vert\; \PP) \land \neg[\tti(\neg[\alpha],\N\to \PP)\;\vert\; \PP]\\
    \textit{Can input on channels in type $\alpha$ and cannot input on $\neg\alpha$, and continues as such.}
\end{array}\]

Native type theory is intended to be a practical method to equip programming languages with a shared system of higher-order reasoning. The authors believe that the potential applications are significant and broad, and we encourage community development.

\subsection{Motivation and implementation}
\label{ssec:motivation}


As software systems become increasingly complex, it is critical to develop adequate frameworks for reasoning about code systematically across languages. By generating type systems for programming languages, native type theory can improve control, reasoning, and communication of systems.

For example, web browsers use the dynamic, weakly-typed language of JavaScript. Companies have recognized that correct and maintainable code requires static type checking. Microsoft's TypeScript \cite{microsoft}, Facebook's Flow \cite{facebook}, and Google's Closure Compiler \cite{google} are multi-million dollar efforts to retrofit JavaScript with a strong, static type system; yet none of these is sound. When presented as a structured $\lambda$-theory \cite{js-k}, JavaScript has a native type system which is sound by construction.

Native type theory is intended to be implemented as a development environment, based on a library of formal semantics and translations, in which one can program in languages enhanced by their native type systems. Code can be written in the same way, but enriched with predicates and dependent types, both (1) to condition existing codebases and (2) to expand software capability.

To this end, we plan to leverage progress in language specification. K Framework \cite{kframe} is a formal verification tool which is used to give complete semantics of many popular languages, including JavaScript, C, Java, Python, Haskell, LLVM, Solidity, and more.  These specifications can be presented as \textit{$\lambda$-theories with equality} ($\S$\ref{sec:lamthy}), and input to native type theory.

The type system generated can then be used for many purposes, e.g.\ to query codebases. The search engine Hoogle \cite{hoogle} queries Haskell libraries by function signature. This idea can be expanded to many languages and strengthened by more expressive types. If $\varphi:\tts\to \Prop$ is a predicate on $\tts$-terms and $\psi:\ttt\to \Prop$ is one on $\ttt$-terms, e.g.\ a security property, we can form the type of programs $\tts\to \ttt$ for which substituting $\varphi$ entails $\psi$ ($\S$\ref{ssec:sub}, def.\ \ref{eq:hom}).
\[
    [\varphi, \psi] := \{\lambda x.c:\tts\to \ttt \;|\; \forall p : \tts.\; \varphi(p) \ra \psi(c[p/x])\}
\]

Of course, the full applications of native type systems require substantial development. Most basic is the need for efficient type-checking, but this is well-studied \cite{coqTypeCheck}. For usability, there will need to be libraries of native types, so programmers can express useful ideas without overly complex formulae.

The larger endeavor, to create a framework for reasoning across many languages, calls for developing a public library of both formal semantics and translations between languages.

\subsection{Organization and contribution}

Our goal is to demonstrate that composing two categorical ideas can be highly useful to computer science. In the process we emphasize many ideas that may be ``known'' in theory but are not widely known nor used in practice. The main original contribution is that by internalizing transition systems in $\lambda$-theories, native type systems can reason about both the structure and behavior of terms simultaneously.

$\S$\ref{sec:lamthy} \textbf{Structured $\lambda$-theories}.
We define \textit{$\lambda$-theories with equality} as cartesian closed categories with pullbacks, and we interpret the internal language as simply-typed $\lambda$-calculus combined with the syntax of generalized algebraic theories \cite{gats}.

Theories ordinarily model the structure of programs, while behavior is modelled separately \cite{opsem}; but in fact transition systems can be modelled as internal categories. The concept of ``language equipped with a notion of behavior'' motivates the 2-category of \textit{structured $\lambda$-theories}. We define the $\uprho\pi$-calculus \cite{rhocal}, a concurrent language with reflection, as our running example for native types.

$\S$\ref{sec:topos} \textbf{Logic in a presheaf topos}. A $\lambda$-theory $\T$ embeds into a presheaf topos $\pow(\T)$, and we develop its internal language. Predicates on the sorts of $\T$ form a $\lambda$-theory $\omega\T$ which refines the entire language; refined binding is then applied to condition program input ($\S$\ref{ssec:hom}).

We show that the predicate and codomain fibrations of $\pow(\T)$ form a ``cosmic'' \textit{higher-order dependent type theory} (HDT), and this construction is 2-functorial.

Hence \textit{native type theory} is the composite 2-functor
\[\begin{tikzcd}
	{\lambda\mathrm{Thy}_{=}^{\mathsf{op}}} & {\mathrm{Topos}} & {\mathrm{HDT}\Sigma.}
	\arrow["{\mathscr{P}}", from=1-1, to=1-2]
	\arrow["{\mathscr{L}}", from=1-2, to=1-3]
\end{tikzcd}\]
This extends to structured $\lambda$-theories, i.e.\ the arrow 2-categories over this composite. 


$\S$\ref{sec:typeth} \textbf{Native type theory}. The native type system of a $\lambda$-theory $\T$ is presented as the internal language of the presheaf topos, $\msc{L}\pow(\T)$. The system is an extension of \textit{higher-order dependent type theory} \cite{jacobs}, as in the Calculus of Constructions \cite{coc}. We present the system as generated by $\T$, and give the rules for types and terms, as well as those for functoriality.

$\S$\ref{sec:apply} \textbf{Applications}.
We explore a few kinds of applications: conditioning term behavior, with subgraphs of rewrite systems and modalities, and deriving behavioral equivalence; conditioning program input with refined binding, and reasoning about contexts with predicate homs; and translating types across programming paradigms.
The scope of applications is beyond what can be given here.

\section{Structured lambda-theories}
\label{sec:lamthy}

Simply-typed $\lambda$-calculus is the language of products and functions. It is regarded as the foundation of computer science \cite{barendregt} and much of modern programming \cite{harper}.

The syntax of a language can be modelled by a \textit{syntactic category}, in which an object is a sorted variable context, a morphism is a term constructor, and composition is substitution. The $\lambda$-calculus is the syntax of \textit{cartesian closed categories} \cite{lambek}.

A particular $\lambda$-calculus or \textit{$\lambda$-theory} is presented by sorts, constructors, and equations. This is just like presenting an algebraic structure such as groups, but with higher-order constructors (ex.\ \ref{ssec:rho}, $\msf{input}$). References for the syntax and semantics of simply-typed $\lambda$-calculus are \cite[Ch.\ 4]{crole} and \cite[Ch.\ 2]{jacobs}. To save space in presentations, we denote products by $\tts,\ttt$ and functions by $[\tts\to \ttt]$.

The main rules of a $\lambda$-theory define how to construct and use functions.
\[\begin{prooftree}
    \hypo{\Gamma, x'\tts \vdash t:\ttt}
    \infer1[abstraction]{\Gamma \vdash \lambda x.t:[\tts\to \ttt]}
\end{prooftree} \quad \quad \quad \quad \quad \quad \quad \quad
\begin{prooftree}
    \hypo{\Gamma \vdash \lambda x.t:[\tts\to \ttt], u:\tts}
    \infer1[application]{\Gamma \vdash t[u/x]:\ttt}
\end{prooftree}\]

\begin{definition}
    A \define{$\lambda$-theory with equality} is a cartesian closed category with pullbacks, also known as a ``properly cartesian closed category'' \cite{elephant}. The 2-category of $\lambda$-theories with equality, finitely continuous closed functors, and natural transformations is $\lambda\Thy_{=}.$
\end{definition}    

The syntax of a $\lambda$-theory with equality can be derived from its subobject fibration having fibered equality \cite[Ch.\ 3]{jacobs}. We interpret the language as simply-typed $\lambda$-calculus combined with the syntax of \textit{generalized algebraic theories} \cite{gats}, which provide \textit{indexed sorts}.

\begin{minipage}{0.2\textwidth}
\[\begin{prooftree}
    \hypo{\Gamma \vdash x_1:\tts_1,\dots,x_n:\tts_n}
    \infer1[sort symbol]
    {\Gamma, \vec{x_i}:\vec{\tts_i}\vdash \mtt{A}(x_1,\dots,x_n) \;\; \msf{sort}}
\end{prooftree}\]
\end{minipage} \qquad \qquad \qquad \qquad \qquad \qquad
\begin{minipage}{0.2\textwidth}
\[\begin{prooftree}
    \hypo{\Gamma \vdash s_1:\tts_1, \dots, s_n:\tts_n}
    \infer1[term symbol]
    {\Gamma \vdash \ttf(s_1,\dots,s_n):\tts}
\end{prooftree}\]
\end{minipage}

\vspace{1em}

Indexed sorts are highly expressive. If we take the source map of a graph $s:\E\to \V$ as an indexed sort, then $s(v)$ is the sort of edges out of a vertex --- i.e., the behavior of a term.


Henceforth, ``$\lambda$-theory'' means $\lambda$-theory with equality.

\subsection*{$\lambda$-theories with behavior}

What $\lambda$-theories do not explicitly represent is the \textit{process} of computation. In practice, computing consists not of equations but \textit{transitions}. There are many ways to model the behavior of languages \cite{opsem}, but the operational semantics of higher-order languages is still in development \cite{hirsch}. We introduce a method of representing behavior internally.

A language with a rewrite system can be modelled by a $\lambda$-theory $\T$ with an internal category, which includes constructors and equations to specify the interaction between rewrites and constructors, a.k.a.\ the \textit{operational semantics}. First, here is the theory of categories.

\begin{definition} $\msf{Th.Cat}$
\[\begin{array}{rlcrlcrcl}
    \mtt{Hom}: & \E\to \V,\V &&
    ;_{abc}: & \Hom(a,b),\Hom(b,c)\to \Hom(a,c) && (e_1;e_2);e_3 & = & e_1;(e_2;e_3)
    \\
    &&& \mtt{id}_a: & 1\to \Hom(a,a) && \mtt{id}_a;e = e && e;\mtt{id}_b = e
\end{array}\]
\end{definition}
Given $(a,b):\Gamma\to \V,\V$ we denote $e:\Gamma\to \Hom(a,b)$ by $e(\vec{x}): a(\vec{x}) \leadsto b(\vec{x}).$

Note --- Though composition is useful, we often want to reason about ``basic rewrites'' or single-step computations. For most of the paper we will simply use an internal graph. It is easy to combine both approaches, by distinguishing one sort for edges and one sort for morphisms.

Operational semantics describes how term constructors interact with the transition system \cite{opsem}: given a constructor $\ttf:\prod \tts_i\to \ttt$ and terms $v_i:\tts_i$ with edges $e_{ij}: v_i \leadsto w_{ij}$, what is the behavior of $\ttf(v_1,\dots, v_n)$?
\[\begin{prooftree}
    \hypo{\{e_{1j}: v_1 \leadsto w_{1j}\}}
    \hypo{\dots}
    \hypo{\{e_{nj}:v_n\leadsto w_{nj}\}}
    \infer3[$\msf{R}(\ttf)$]{\{e_k:\ttf(v_1,\dots,v_n)\leadsto \ttg_k\}}
\end{prooftree}\]

To specify this interaction, we take the source map $s:\E\to \tts$ as an indexed sort $\tts^\ast(x)$; then $\tts^\ast(v)$ are the edges with source $v$. This allows us to define operational semantics in a $\lambda$-theory.

\begin{definition}
\label{def:opsem}
    Let $\T$ be a $\lambda$-theory with sorts $\tts_i$, constructors $\ttf_{ij}:\prod_j \tts_{ij}\to \tts_i$, and graphs for each sort.
    A \define{behavior rule} for a term constructor $\ttf:\prod \tts_i\to \tts$ is a constructor
    \[\msf{R}(\ttf)_{\vec{v}}:\prod \tts^\ast_i(v_i)\to \tts^\ast(\ttf(\vec{v}))\]
    such that $\msf{R}(\ttf)_{\vec{v}}(e_1, \dots, e_n): \ttf(\vec{v}) \leadsto \ttg(\vec{e})$, where $\ttg:\prod \tts^\ast_i(v_i)\to \tts$.
    
    An \define{operational semantics} for $\T$ is for each sort a family of edges $\{\mtt{r}_{ij}(\vec{x}):\mtt{a}_{ij}(\vec{x})\leadsto \mtt{b}_{ij}(\vec{x}):\E_{\tts_i}\}$ and for some term constructors a behavior rule $\{\langle \ttf_{ij},\msf{R}(\ttf_{ij})\rangle\}$. This defines a subtheory $\msf{O}(\T)\to \T$.
\end{definition}

\begin{theorem}
    Behavior rules correspond to GSOS rules \cite{opsem} for deterministic labelled transition systems. The general case can be derived using an internal relation on $\V,\mtt{A},\V$, where $\mtt{A}$ is a sort of actions.
\end{theorem}

Hence, internal operational semantics are equivalent to GSOS distributive laws. Providing a novel perspective to a basic topic in computer science, this connection warrants exploration in future work.


By representing behavior internally, we will see that
\begin{center}
    \textit{native type systems reason about both the structure and behavior of terms}.
\end{center}
For example, there can be a predicate for ``contexts $\lambda x.c: \tts\to \ttt$ such that if $a:\tts$ satisfies $\varphi$ then for all $e_i:c[a/x]\leadsto b$ if $\psi_i(b)$ then no step of $e$ satisfies $\epsilon$''. Combining both kinds of reasoning is extremely expressive, and the applications in $\S$\ref{sec:apply} provide only a modest glimpse.

\begin{example}
\label{ssec:rho}
$\uprho\pi$-calculus $\msf{Th}.\uprho\pi$ (polyadic)

The $\uprho\pi$-calculus or \textbf{r}eflective \textbf{h}igher-\textbf{o}rder $\pi$-calculus \cite{rhocal} is a concurrent language succeeding the $\pi$-calculus \cite{polyadic}. It is the language of the blockchain platform RChain \cite{rchain}.

The $\uprho\pi$-calculus has sorts $\PP$ and $\N$ for processes and names, which act as code and data respectively. Reference $@$ and execute $\ast$ transform one into the other. Terms are built up from the null process $\ttz$ by parallel $-\vert -$, output $\tto$, and input $\tti$. The basic rule is $\mtt{comm}$: an output and input process in parallel on the same name can \textit{communicate}, transfering a list of processes as data.

\[\begin{array}{rlcrlcrl}
    \ttz: & 1\to \PP && 
    -\vert-: & \PP,\PP \to \PP &&
    (\PP,-|-,\ttz) & \text{commutative monoid}\\
    @: & \PP \to \N && 
    \mtt{out}_k: & \N,\PP^k \to \PP && 
    \mtt{run}: & \PP\to \E\\
    \ast: & \N \to \PP &&
    \mtt{in}_k: & \N,[\N^k\to \PP]\to \PP &&
    \mtt{comm}_k: & \N,\PP^k,[\N^k\to \PP] \to \E
\end{array}\]
\[\begin{array}{rcl}
    \mtt{comm}_k(n,\vec{q_i},\lambda \vec{x_i}.p) & : & \tto(n,\vec{q_i}) \;|\; \tti(n,\lambda \vec{x_i}.p) \leadsto p[@q_i/x_i] \\
    \mtt{run}(p) & : & \ast(@p) \leadsto p
\end{array}\]
\[\begin{array}{lclcl}
    (s,t):\E\to \PP,\PP &&
    \mtt{par}_l: \E,\PP \to \E &&
    \mtt{par}_l(\mtt{r} , q) : s(\mtt{r})\vert q \leadsto t(\mtt{r})\vert q \\
    \mtt{par}_r: \PP, \E \to \E &&
    \mtt{par}_r(p,\mtt{r}) = \mtt{par}_l(\mtt{r},p) &&
    \mtt{par}_l \;\;\; \text{c. monoid action of }\PP\mbox{ on }\E
\end{array}\]

The $\uprho\pi$-calculus is our running example of a $\lambda$-theory. In the native type system $\msc{L}\pow(\T)$  ($\S$\ref{sec:typeth}) of $\T = \msf{Th}.\uprho\pi$, a predicate on names $\alpha: y\N\to \Prop$ is called a \textit{namespace} \cite{namespace}, and a predicate on processes $\varphi: y\PP\to \Prop$ is called a \textit{codespace}.
\end{example}

In distinguishing internal graphs and edge constructors for the behavior of terms, we should also require that morphisms of ``theories with behavior'' should \textit{respect} this structure. We generalize to define ``structure'' as any $\lambda$-theory morphism into $\T$.
\begin{definition}
    A \define{structured $\lambda$-theory} is a $\lambda$-theory with equality $\T$ equipped with a morphism $\tau:\ess\to \T$. The 2-category of $\ess$-structured $\lambda$-theories is the strict coslice 2-category $\ess/\lambda\Thy_=$. The 2-category of all structured $\lambda$-theories is the strict arrow 2-category $[\mrm{I},\lambda\Thy_=]$.
\end{definition}


Because native type theory is functorial, a structure $\tau:\ess\to \T$ translates types of $\T$ into types of $\ess$. For including behavior, this simply distinguishes the ``behavioral'' types; for more complex structures, the translation may be highly expressive. As the concept is very general, we give a few more examples.


\begin{example} \textbf{Sorting}
    A polyadic language such as the $\uprho\pi$-calculus can be refined with many sorts of name, and name sorts given sorted arities for input and output. This is used by Milner \cite{polyadic} to designate channels to send and receive certain kinds of data.
    
    Sorting is a structure $\sigma: S(\msf{Th}.\uprho\pi)\to \msf{Th}.\uprho\pi$, where the fiber over $\N$ is the set of name sorts, and the fibers over $\tti_k$ and $\tto_k$ are the sorted inputs and outputs with total arity $k$.
    
    The native type system of the structured $\lambda$-theory contains and converts between the sorted and unsorted language. By functoriality, the type system expands to the 2-category of all sorted $\uprho\pi$-calculi.
\end{example}

\begin{example} \textbf{Encoding}
    Because a structure is simply a morphism of $\lambda$-theories, we can consider any translation $\tau:\ess\to \T$ as a structured $\lambda$-theory. This can be understood as compiling or encoding the programs of $\ess$ into those of $\T$.
    
    For example if we encode a complex language like C++ into a simpler language like name-passing $\lambda$-calculus, then we can identify a type of $\lambda$-term with good properties and take its preimage along $\tau$ to form a type of well-behaved C++ programs.
\end{example}

Another example is mapping a theory into a computing environment, such as a virtual machine. This allows for reasoning about both languages and their implementation.





From a structured $\lambda$-theory we derive a native type system, using the presheaf construction, and demonstrate how it can be used to reason about the structure and behavior of terms.


\section{The Logic of a Presheaf Topos}
\label{sec:topos}

Topos theory \cite{sheavesinGL} expands the domain of predicate logic and intuitionistic type theory \cite{martin-lof} beyond sets and functions. Most useful is the fact that every category embeds into a topos. For any $\lambda$-theory, the internal language of its presheaf topos is its native type system.

Let $\T$ be a $\lambda$-theory. The category of \textit{presheaves} is the functor category $[\T^\op,\Set]$, denoted $\pow(\T)$. This defines a 2-functor to elementary toposes and geometric morphisms
\[\pow:\lambda\Thy_=^\op\to \Topos \quad \quad \pow(\F) = (\mrm{Lan}_\F\dashv \F^\ast): [\T^\op,\Set]\to [\ess^\op,\Set].\]
where $\mrm{Lan}_\F$ is left Kan extension \cite[Ch.\ 10]{working} and $\F^\ast$ is precomposition by $\F:\ess\to \T$.

A presheaf is a context-indexed set of data on the sorts of a theory. The canonical example is a \textit{representable} presheaf, of the form $\T(-,\tts)$, which indexes all terms of sort $\tts$. The Yoneda embedding $y:\T\to \pow(\T)::\tts\mapsto \T(-,\tts)$ preserves limits and internal homs.


    
A \textit{subobject classifier} is an object $\Omega$ with a natural isomorphism $\mrm{c}:\T(-,\Omega)\simeq \Sub(-):\T^\op\to\Pos.$ We may denote $\Omega$ as $\Prop$; this is its role in the type system: a \textit{predicate} is a morphism $\varphi:A\to \Omega$, and the \textit{comprehension} of $\varphi$ is the subobject $\mrm{c}(\varphi) \; := \; \{a:A \;|\; \varphi(a)\} \rtail A.$

A \define{topos} is a $\lambda$-theory with equality with a subobject classifier. For presheaves, the hom and subobject classifier are defined $[P,Q](\tts) = \pow(\T)(y(\tts)\times P,Q)$ and $\Omega(\tts) = \{\varphi\rtail y(\tts)\}.$

The values of $\Omega$ can be understood as $\Omega(\tts) \simeq \{\text{sieves on } \tts\}.$
A \textbf{sieve} on $\tts$ is a set of morphisms into $\tts$ closed under precomposition. A simple example is a \textit{principal sieve} $\langle \ttf\rangle:\Omega(\ttt)$ generated by $\ttf:\tts\to\ttt$.
\[\langle \ttf\rangle(\ttr) = \{ t:\ttr\to \ttt \;|\; \exists u'\ttr\to \tts.\; \ttf(u) = t\} \] 

A sieve can be understood as a set of shapes of abstract syntax tree with (sorted) holes for leaves, closed under substitution. These are the basic objects of reasoning in native type theory, as they are predicates on representable presheaves.

\begin{example}
\label{eq:sieve}
    In the $\uprho\pi$-calculus (ex.\ \ref{ssec:rho}) we can define a context $c(n):\PP\to \PP$ which replicates processes on a name $n:1\to \N$.
    \[\begin{array}{llcll}
    c(n)& := \tti(n, \lambda x.\{\tto(n,\ast x) \; \vert \ast x \} ) &&
    !(-)(n) & := \tto(n,\{c(n)\vert -\}) \; \vert \; c(n).
    \end{array}\]
One can check that $!(p)(n) \;\; \leadsto \;\; !(p)(n) \; \vert \; p$ for any process $p$. The sieve $\langle!(-)(n)\rangle:\Omega(\PP)$ consists of processes which replicate on the name $n$ by the above method.
\end{example}

For simpler formulae, we denote the values of a presheaf by $A_\tts := A(\tts)$, and the action of $u:\ttr\to \tts$ by $-\cdot u:= A(u): A(\tts)\to A(\ttr)$. For $\varphi:A\to \Prop$ we denote $\varphi_\tts^a := \varphi(\tts)(a)$; more generally for any $p:P\to A$ we denote $p_\tts^a:=p_\tts^{-1}(a)$ as the \textit{fiber} over $a$ ($\S$\ref{ssec:cod}). Finally, sorts and term constructors are identified with their images under $y$, so $\tts$ will mean $y\tts$ when applicable.

\subsection{The predicate fibration}
\label{ssec:sub}

For any $\lambda$-theory $\T$, there is a ``category of predicates'' $\Omega\pow(\T)$ over $\pow(\T)$ where the fiber over each presheaf is its poset of predicates. Quantification gives adjoints to change-of-base between fibers; we show that moreover the domain is cartesian closed, complete and cocomplete. The structure of this fibration provides higher-order predicate logic of the presheaves on $\T$.

We use $\Omega^A$ to denote the poset of predicates $\varphi:A\to \Omega$, ordered by entailment. The \textit{predicate functor} of $\pow(\T)$ is $\Omega^{(-)}:\pow(\T)^\op\to \Pos.$ For $f:A\to B$, precomposition of predicates corresponds to preimage of subobjects. This is written as substitution $\varphi[f] := \Omega^f(\varphi)$. 

Substitution can be understood as \textit{pattern-matching}.

\begin{example}
    For a $\uprho\pi$-calculus predicate $\varphi:y(\PP)\to \Prop$, substitution by $y(\tti):y(\N)\times y([\N,\PP])\to y(\PP)$ is the query ``inputting on what name-context pairs yield property $\varphi$?''
    \[\varphi[\tti]_\tts = \{\tts\vdash (n,\lambda x.p):\N, [\N\to \PP] \;|\; \varphi(\tti(n,\lambda x.p))\}\]
\end{example}

Each poset $\Omega^A$ is in fact a complete Heyting algebra: meet and join are intersection and union, $\top = A$ and $\bot = (\tts\mapsto \emptyset)$, implication is 
\[(\varphi\ra \psi)_\tts(a) :=\;\; \Pi u'\ttr\to\tts.\; \varphi_\ttr(a\cdot u)\ra \psi_\ttr(a\cdot u)\]
and negation is $\neg[\varphi]:= (\varphi \ra \bot$).

We can assemble the image of $\Omega^{(-)}$ into one category, with the Grothendieck construction \cite[1.10]{jacobs}.

\begin{definition}
The \emph{category of predicates} of $\pow(\T)$, denoted $\Omega\pow(\T)$, is defined as follows.
\[\begin{array}{l@{\;\;}c@{\;\;}l}
    \text{Object} & \text{a pair } & \langle A \;,\; \varphi:\Omega^A\rangle \\
    \text{Morphism} & \text{a pair } & f:\langle A,\varphi\rangle \to \langle B,\psi\rangle\\
    & = & \langle f:A\to B \;,\; \varphi\ra \psi[f]\rangle\\
    \text{Composition} && f;g : \langle A,\varphi\rangle \to \langle B,\psi\rangle\to \langle C,\chi\rangle\\ & = & \langle f;g:A\to C, \varphi\ra \psi[f]\ra \chi[g][f]\rangle
\end{array}\]
The projection $\pi_\Omega:\Omega\pow(\T)\to \pow(\T)$ is the \define{predicate fibration}; the fiber over $A$ is $\Omega^A$, and the fiber over $f:A\to B$ is $\Omega^f:\Omega^B\to \Omega^A$, known as a \emph{change-of-base functor}.
\end{definition}


A fibration is a functor with a well-behaved notion of preimage, used in type theory for \textit{indexing}; a reference is \cite[Ch.\ 1]{jacobs}. The predicate fibration is highly structured: each change-of-base functor has adjoints which are \textit{dependent sum} and \textit{dependent product}.


\begin{proposition}
The projection $\pi_\Omega:\Omega\pow(\T)\to \pow(\T)$ has \define{indexed sums and products} \cite{jacobs}: for each $f:A\to B$, the functor $\Omega^f:\Omega^B\to \Omega^A$ has left and right adjoints $\exists_f\dashv \Omega^f\dashv \forall_f$.
\[\begin{array}{rlcrl}
    \exists_f(\varphi)_\tts^b := & \exists a'A_\tts.\; (f_\tts(a)=b) \land \varphi(a) && 
    \forall_f(\varphi)_\tts^b := & \forall(u'\ttr\to \tts).\forall a'A.\; (f_\ttr(a)=b) \ra \varphi(a)
\end{array}\]

 \end{proposition}

The left adjoint $\exists_f$ is called \define{direct image}, because on subobjects it is composition by $f$; we call the right adjoint $\forall_f$ \define{secure image}. While $\Omega^f$ is a morphism of complete Heyting algebras, $\exists_f$ and $\forall_f$ are only morphisms of join and meet semilattices, respectively.


\begin{example}
\label{ex:step}
    Let $\msf{Th.Gph}\to \T$ be a $\lambda$-theory with a graph, and $\varphi:\mtt{V}\to \Prop$ be a predicate on terms. Then $\varphi[s]:\E\to \Prop$ are rewrites with $\varphi($source$)$, and $\exists_{t}(\varphi[s])$ are the targets of these rewrites. Hence there is a \textit{step-forward} operation $\F_! := [s];\exists_t: [\V,\Prop]\to [\V,\Prop]$.
    
    The \textit{secure step-forward} is a more refined operation: $\F_*(\varphi):=\forall_{t}(\varphi[s])$ determines the terms $u$ for which $(t\leadsto u)\ra \varphi(t)$. For security protocols, this can filter agents by past behavior.
\end{example}


The change-of-base adjoints satisfy the \textit{Beck--Chevalley condition}: this means that quantification commutes with substitution, and implies that $\Omega^{(-)}:\pow(\T)^\op\to \CHA$ is a \textit{first-order hyperdoctrine} \cite{hyper} and a \define{higher-order fibration} \cite[section 5.3]{jacobs}.


This concept leaves implicit additional structure: there is also an \textit{internal hom} of predicates.

\begin{proposition}
\label{eq:hom}
    $\Omega\pow(\T)$ is cartesian closed, as is $\pi_\Omega$. Let $\varphi:A\to \Prop$, $\psi:B\to \Prop$, and let $\langle\pi_1, \pi_2,ev\rangle:A\times [A,B]\to A\times [A,B]\times B.$
    Then $[\varphi, \psi]:[A,B]\to \Prop$ is defined $[\varphi,\psi]:=\forall_{\pi_2}(\varphi[\pi_1]\ra \psi[ev])$. This determines maps $f:A\to B$ for which $\varphi(a)\ra \psi(f(a))$.
\end{proposition}

The cartesian closed structure of $\Omega\pow(\T)$ is significant, because the category of predicates on $\T$ is itself a $\lambda$-theory, the refinement of the language. We explore applications in $\S$\ref{ssec:hom}.

\begin{definition}
\label{def:predthy}
    The \define{predicate theory} of $\T$, denoted $\omega\T$, is the pullback of the predicate fibration along the embedding $y:\T\to \pow(\T)$; it is a $\lambda$-theory fibered over $\T$.
\end{definition}

Note --- We emphasize the idea of having ``lifted'' the language by an abuse of notation: for any operation $\ttf:\tts\to \ttt$, we may denote $\exists_{y(\ttf)}:[y(\tts),\Prop]\to [y(\ttt),\Prop]$ simply by $\ttf$, and $\forall_{y(\ttf)}$ by $\ttf_*$. 

\begin{example}
    As an example of contexts which ensure implications across substitution, we can construct the ``magic wand'' of separation logic \cite{refine}. Let $\T_h$ be the theory of a commutative monoid $(H,\cup,e)$, plus constructors for the elements of a heap. If we define $(\varphi \wand \psi) := [\varphi, \psi][\lambda x.x\cup -]$, then $(\varphi \wand \psi)(h_1)$ means that $\varphi(h_2)\ra \psi(h_1\cup h_2)$.
\end{example}

There is a more expressive way to form hom predicates, which provides \textit{predicate binding}.

\begin{proposition}
\label{eq:reify}
    Let $A,B:\pow(\T)$, and let $\msf{L}_{A,B}:[[A,B],\Prop]\to [[A,\Prop],[B,\Prop]]$ be curried evaluation. There is a right adjoint which we call \define{reification}.
    The predicate $\msf{R}_{A,B}(F)$, denoted $\chi.F$, determines $f:[A,B]$ whose images are contained in those of $F$:
    \[[\chi.F]_\tts(f) \; =\;  \Pi \chi'[A\to \Prop].\;  \exists_f(y(\tts)\times \chi)\ra F(\chi).\]
\end{proposition}

The authors do not know of existing literature on this right adjoint; we do not yet how it is connected with dependent products of $\pow(\T)$. We know it is highly expressive: using reification, separation logic can be generalized from pairs  to \textit{functions of predicates}. We do not know if this has been studied.
    
In addition, the category of predicates has all limits and colimits, by a result of \cite{burstall}. These can be used to form modalities, inductive and coinductive types, and more.
    
\begin{proposition}
    $\Omega\pow(\T)$ is complete and cocomplete, and $\pi_\Omega$ preserves limits and colimits. They are computed pointwise; letting $\pi,\iota$ represent the cone and cocone:
    \[\mrm{lim}_i\langle A_i,\varphi_i\rangle = \langle \mrm{lim}_i(A_i),\mrm{lim}_i(\Omega^{\pi_i}\varphi_i)\rangle
    \quad\quad
    \mrm{colim}_i\langle A_i,\varphi_i\rangle = \langle \mrm{colim}_i(A_i),\mrm{colim}_i(\Sigma_{\iota_i}\varphi_i)\rangle.
    \]
\end{proposition}

    

    
 
To summarize the rich structure present, we allude to a term from category theory: a \textit{cosmos} is a monoidal closed category which is complete and cocomplete \cite{cosmos}.

\begin{proposition}
The predicate fibration $\pi_\Omega:\Omega\pow(\T)\to \pow(\T)$ is a higher-order fibration which is \define{cosmic}: cartesian closed, complete and cocomplete.
\end{proposition}

\subsection{The codomain fibration}
\label{ssec:cod}

Predicates $\varphi:A\to \Prop$ correspond to subobjects $\mrm{c}(\varphi)\rtail A$. More generally, any $p:P\to A$ can be understood as a \textit{dependent type}. This generalizes subsets to indexed sets: fibers over $A$ are expanded from truth values to sets, denoted $p_\tts^a$ or $P_\tts[a]$, and fibers over $\pow(\T)$ are expanded from posets to categories.

Each term constructor $\ttf:\tts\to\ttt$ defines a dependent type $y(\ttf):y(\tts)\to y(\ttt)$, still denoted $\ttf$. Its terms are like the principal sieve $\langle \ttf\rangle:y(\ttt)\to \Prop$, except that the substituted terms $u$ are recorded.
\[\ttf_\ttr[t] = \Sigma u'\ttr\to \tts. \;(f(u) = t)\]

\begin{proposition}
    Let $\mrm{CCT}$ be the category of complete and cocomplete toposes and logical functors. There is a functor $\Delta: \pow(\T)^\op\to \mrm{CCT}$ that maps $A$ to $\pow(\T)/A$ and $f:A\to B$ to pullback.
\end{proposition}

We can denote pullback by substitution, $p[f]_\tts^a := \Delta^f(p)_\tts^a = p_\tts^{f_\tts(a)} = p_\tts^{-1}(f_\tts(a))$.
Dependent sum $\Sigma_f$ and dependent product $\Pi_f$ have the same formulae as predicates, and they commute with substitution.
\[\begin{array}{rlcrl}
    \Sigma_f(\varphi)_\tts^b := & \Sigma a'A_\tts.\; (f_\tts(a)=b) \land \varphi(a) && 
    \Pi_f(\varphi)_\tts^b := & \Pi(u'\ttr\to \tts).\Pi a'A.\; (f_\ttr(a)=b) \ra \varphi(a)
\end{array}\]
The Grothendieck construction of $\Delta$ determines a category over $\pow(\T)$.

\begin{definition}
The \emph{category of dependent types} of $\pow(\T)$, denoted $\Delta\pow(\T)$, is equivalent to the arrow category of $\pow(\T)$. The \define{codomain fibration} is the projection
$\pi_\Delta:\Delta\pow(\T)\to \pow(\T).$
\end{definition}

\begin{proposition}
The codomain fibration $\pi_\Delta$ is a \textit{closed comprehension category} \cite[Sec.\ 10.5]{jacobs} which is cosmic, i.e. cartesian closed, complete and cocomplete.
\end{proposition}

The two fibrations are connected by an adjunction: comprehension interprets a predicate as a dependent type, and factorization takes a dependent type to its image predicate. This fibered adjunction is a \textbf{higher-order dependent type theory} \cite[Sec.\ 11.6]{jacobs}. 
\[\begin{tikzcd}
    	{\Omega\pow(\T)} && {\Delta\pow(\T)} \\
    	& {\pow(\T)}
	\arrow["{\mrm{c}}"{name=0, description}, from=1-1, to=1-3, shift right=3, tail]
	\arrow["\mrm{i}"{name=1, description}, from=1-3, to=1-1, shift right=3]
	\arrow["{\pi_\Delta}", from=1-3, to=2-2, shift left=3]
	\arrow["{\pi_\Omega}"', from=1-1, to=2-2, shift right=3]
	\arrow["\dashv"{rotate=-90}, from=1, to=0, phantom]
    \end{tikzcd}\]
We denote by $\mrm{HDT}\Sigma$ the 2-category of higher-order dependent type theories, a full sub-2-category of adjunctions in the 2-category of fibrations. 

The reason for the notation $\mrm{HDT}\Sigma$ is as follows. Recall that the 2-functor
\[\pow:\lambda\Thy_=^\op\to \Topos\]
sends $\F:\ess\to \T$ to precomposition by $\F$, $\pow(\F):= -\circ \F:[\T^\op,\Set]\to [\ess^\op,\Set]$. This functor preserves pullbacks, inducing morphisms of predicate and codomain fibrations; but it is not locally cartesian closed, nor does it preserve the subobject classifier. It is future work to consider theory translations for which $\pow(\F)$ preserves $\Pi$ and $\Omega$ \cite[C 3]{elephant}.

\begin{theorem}
    The construction which sends a topos to its \define{internal language} $\msc{L}(\msc{E})=\langle \pi_{\Omega\msc{E}},\pi_{\Delta\msc{E}},\mrm{i}_\msc{E},\mrm{c}_\msc{E}\rangle$, consisting of the predicate and codomain fibrations connected by the image-comprehension adjunction, defines a 2-functor $\msc{L}:\Topos\to \mrm{HDT}\Sigma$.
\end{theorem}



\section{Native Type Theory}
\label{sec:typeth}

We present the \define{native type system} $\msc{L}\pow(\T)$ of a $\lambda$-theory with equality $\T$ ($\S$\ref{sec:lamthy}). As $y:\T\to \pow(\T)$ is full and faithful, $\msc{L}\pow(\T)$ is a conservative extension of $\T$.

The system is \textit{higher-order dependent type theory} \cite[Sec.\ 11.5]{jacobs} ``parameterized'' by $\T$. We do not present Equality and Quotient types. We encode Subtyping, Hom, Reification, and Inductive types, which we use in applications.

The type system has \define{predicates} $\msf{x'\Gamma\vdash \varphi:\Prop}$ and \define{types} $\msf{x'\Gamma\vdash A:\Type}$, interpreted as $\varphi:\msf{\Gamma}\to \Omega$ and $\msf{p:A\to \Gamma}$. A term judgement is of the form $\msf{x'\Gamma,a'A\vdash N:B[M]}$, interpreted as a morphism $\msf{\langle M,N\rangle: (A\to \Gamma)\to (B\to \Delta)}$ in the total category of the codomain fibration.

For details on the semantic interpretation of the type system, in particular handling coherence when interpreting substitution as pullback, see Awodey's \textit{natural models} \cite{naturalModels}.

We present the type system as generated from the $\lambda$-theory $\T$, so a programmer can start in the ordinary language and use the ambient logical structure as needed.
\vspace{0.5em}
\begin{itemize}
    \item [$\msf{Y}$] \textbf{Representables} are given in the type system as axioms.
        \[\begin{prooftree}
            \hypo{\interp{\tts:\T}}
            \infer1[$\T_S$]{\msf{y\tts:\Type}}
        \end{prooftree}\quad \quad
        \begin{prooftree}
            \hypo{\interp{\tts_1\vdash \ttf:\tts_2}}
            \infer1[$\T_O$]{\msf{x'y\tts_2 \vdash y\ttf: \Type}}
        \end{prooftree} \quad \quad  \begin{prooftree}
            \hypo{\interp{\tts_1 \vdash \ttf = \ttg:\tts_2}}
            \infer1[$\T_E$]{\msf{x'y\tts_2 \vdash y\ttf = y\ttg}}
        \end{prooftree}\]
    The type $\sfy\tts$ indexes all terms of sort $\tts$. Because the Yoneda embedding preserves limits and internal hom, we have $\sfy(\tts_1,\tts_2) = (\sfy\tts_1,\sfy\tts_2)$ and $\sfy[\tts\to \ttt] = [\sfy\tts\to \sfy\ttt]$.
    
    \item [$\msf{\Sigma}$] \textbf{Dependent Pair} is an indexed sum generalizing existential quantification.
        \[\begin{prooftree}
            \hypo{\msf{\Gamma\vdash A:\Type}}
            \hypo{\msf{\Gamma, x'A\vdash B:\Type}}
            \infer2[$\msf{\Sigma_\FF}$]
            {\msf{\Gamma \vdash \Sigma x'A.B: \Type}}
        \end{prooftree} \quad \quad
        \begin{prooftree}
            \hypo{\msf{\Gamma\vdash a:A}}
            \hypo{\msf{\Gamma \vdash u: B[a/x]}}
            \infer2[$\msf{\Sigma}_\I$]
            {\msf{\Gamma\vdash \langle a,u\rangle :\Sigma x'A.B}}
        \end{prooftree}\]
        \[\begin{prooftree}
            \hypo{\msf{\Gamma, z'\Sigma x'A.B\vdash C:\Type}}
            \hypo{\msf{\Gamma,a'A,u'B\vdash Q:C[\langle a,u\rangle/z]}}
            \infer2[$\msf{\Sigma}_\EE$]
            {\msf{\Gamma, z :\Sigma x'A.B\vdash (z \;\msf{as}\; \langle a,u\rangle \; \msf{in}\; Q):C}}
        \end{prooftree}\]
        \[\begin{array}{lcll}
            \msf{\langle M,N\rangle \;\msf{as}\; \langle a,u\rangle \;\msf{in}\; Q} & = & \msf{Q[M/a,N/u]} & (\msf{\Sigma}_\beta) \\
            \msf{P \;\msf{as}\; \langle a,u\rangle \;\msf{in}\; Q[\langle a,u\rangle/z]} & = & \msf{Q[P/z]} & (\msf{\Sigma}_\eta)
        \end{array}\]
    
    \item [$\msf{\Pi}$] \textbf{Dependent Function} is an indexed product generalizing universal quantification.
        \[\begin{prooftree}
            \hypo{\msf{\Gamma\vdash A: \Type}}
            \hypo{\msf{\Gamma,x'A\vdash B:\Type}}
            \infer2[$\msf{\Pi}_\FF$]
            {\msf{\Gamma \vdash \Pi x'A.B: \Type}}
        \end{prooftree} \quad \quad 
        \begin{prooftree}
                \hypo{\msf{\Gamma, x'A \vdash t:B}}
                \infer1[$\msf{\Pi}_\I$]
                {\msf{\Gamma\vdash \lambda x'A.t :\Pi x'A.B}}
        \end{prooftree}\]
        \[\begin{prooftree}
                \hypo{\msf{\Gamma \vdash f:\Pi x'A.B}}
                \hypo{\msf{\Gamma\vdash u:B}}
                \infer2[$\msf{\Pi}_\EE$]
                {\msf{\Gamma\vdash f(u) :B[u/x]}}
        \end{prooftree} \quad \quad 
        \begin{array}{lcll}
            \msf{(\lambda x'A.t)(a)} & = & \msf{t(a)} & (\msf{\Pi}_\beta) \\
            \msf{f} & = & \msf{\lambda x'A.f} & (\msf{\Pi}_\eta)
        \end{array}\]
We derive existential $\exists$ from $\msf{\Sigma}$ and universal $\forall$ from $\msf{\Pi}$ by image factorization. The rest of predicate logic $\bot,\top,\lor,\land,\ra,\neg$ is also encoded in terms of $\msf{\Sigma}$ and $\msf{\Pi}$.
\vspace{0.5em}            
    \item [$\msf{\{\}}$] \textbf{Comprehension} converts a predicate to the type of its satisfying terms. The rules which convert a type to its image predicate can be derived from $\msf{\Sigma}$ and Equality.
        \[\begin{prooftree}
            \hypo{\msf{\Gamma,x'A\vdash \varphi:\Prop}}
            \infer1[$\msf{c}_\FF$]{\msf{\Gamma \vdash \{x'A \;|\; \varphi\}:\Type}}
        \end{prooftree} \quad \quad \quad 
        \begin{prooftree}
            \hypo{\msf{\Gamma,x'A\vdash \varphi:\Prop}}
            \hypo{\msf{\Gamma\vdash M:A}}
            \hypo{\msf{\Gamma\vdash \varphi[M/x]}}
            \infer3[$\msf{c}_\I$]
            {\msf{\Gamma \vdash \msf{i}(M):\{x'A \;|\; \varphi\}}}
        \end{prooftree}\]
        \[\begin{minipage}{0.2 \textwidth}
        \begin{prooftree}
            \hypo{\msf{\Gamma\vdash N:\{x'A \;|\; \varphi\}}}
            \infer1[$\msf{c}_\EE$]
            {\msf{\Gamma \vdash \msf{o}(N):A}}
        \end{prooftree}
        \end{minipage}\qquad
        \begin{minipage}{0.2 \textwidth}
        $\begin{array}{ll}
            \msf{\msf{o}(\msf{i}(M)) = M} & (\msf{c}_\beta)\\
            \msf{\msf{i}(\msf{o}(N)) = N} & (\msf{c}_\eta)
        \end{array}$
        \end{minipage} \quad \quad 
        \begin{prooftree}
            \hypo{\msf{\Gamma_1, x'A, \Gamma_2,\varphi \vdash \psi}}
            \infer[double]1[$\msf{c}_\EE^\circ$]
            {\msf{\Gamma_1, a:\{x'A \;|\; \varphi\}, \Gamma_2[\msf{o}(a)/x] \vdash \psi[\msf{o}(a)/x]}}
        \end{prooftree}\]


\item [$\subseteq$] \textbf{Subtyping} of predicates is defined $(\varphi\subseteq \psi):= \msf{\forall a'A.\; \varphi(a)\ra \psi(a)}.$

\item[$\to$] \textbf{Hom type} (def. \ref{eq:hom}) of $\msf{A_1\vdash B_1:\Type}$ and $\msf{A_2\vdash B_2:\Type}$ is defined $\msf{\Pi x'A_1}.\;\msf{B_1[\pi]\ra B_2[ev]}$.

\item[$\msf{R}$] \textbf{Reification} (def. \ref{eq:reify}) $\msf{\chi.F:[A,B]\to \Prop}$ is defined $\msf{\Pi \varphi'[A\to \Prop]}.\;\msf{\varphi\ra F(\varphi[-])}$.

\item [$\mu$] \textbf{Inductive type} of $\F:[\msf{A},\Prop]\to [\msf{A},\Prop]$: the least and greatest fixed points are defined 
\[\begin{array}{rcl}
    \msf{\mu \varphi.F(\varphi)} & := & \msf{\exists \varphi'[A,\Prop].\; (\varphi\subseteq F(\varphi))\ra \varphi}\\
    \msf{\nu \varphi.F(\varphi)} & := & \msf{\forall \varphi'[A,\Prop].\; (F(\varphi)\subseteq \varphi)\ra \varphi}
\end{array}\]
These are used to form data structures and modalities; we can generalize to W-types \cite{wtypes}.

\end{itemize}

\vspace{0.5em}

These rules constitute the native type system $\msc{L}\pow(\T)$, abridged for a first presentation. We include rules for functoriality, so that translations of $\lambda$-theories induce translations of native type systems.

\begin{itemize}
\item [$\F$] \textbf{Translation} is given by precomposing types and ``whiskering'' terms.
\[\begin{prooftree}
    \hypo{\interp{\F:\T_1\to \T_2}}
    \hypo{\msf{\Gamma\vdash A:\Type_2}}
    \infer2[$\F_\msf{Ty}$]{\msf{\Gamma\circ F\vdash A\circ F:\Type_1}}
\end{prooftree} \quad \quad \quad 
\begin{prooftree}
    \hypo{\interp{\F:\T_1\to \T_2}}
    \hypo{\msf{x'\Gamma, y'A\vdash N:B[M]}}
    \infer2[$\F_\msf{Tm}$]{\msf{x'(\Gamma\circ F), y'(A\circ F)\vdash N\cdot F:(B\circ F)[M\cdot F]}}
\end{prooftree}\]
We add rules that $\F^\ast: \pow(\T_2)\to \pow(\T_1)$ preserves substitution, dependent pair, and co/limits.
\end{itemize}

To further research we leave the rules for the colax preservation of $\msf{\Pi}$ and $\msf{Prop}$, and the rules for the two covariant functors $\mrm{Lan}_\F,\mrm{Ran}_\F:\pow(\T_1)\to \pow(\T_2)$ given by left and right Kan extension.

As a small demonstration of the type system, suppose we have a program $\ttf:\tts\to \ttt$, and we want to construct the predicate which checks whether a term of sort $\ttt$ has been processed by $\ttf$.
\[\begin{prooftree}
    \hypo{\msf{y\ttt\vdash y\ttf:\Type}}
    \hypo{\msf{y\ttt,y\ttf\vdash y\tts:\Type}}
    \infer2[]{\msf{y\ttt\vdash \Sigma \ttg'y\ttf.y\tts:\Type}}
\end{prooftree} \quad \quad \quad \quad 
\begin{prooftree}
    \hypo{\msf{y\ttt\vdash \ttg:y\ttf}}
    \hypo{\msf{y\ttt, x'y\ttf\vdash u:y\tts[\ttg/x]}}
    \infer2[]{\msf{y\ttt\vdash \langle \ttg,u\rangle:\Sigma \ttg'y\ttf.y\tts.}}
\end{prooftree}\]
We can then write protocols based on this precondition in the native type system.


\section{Applications}
\label{sec:apply}

Native type systems are highly expressive and versatile. We demonstrate a few small examples. Notation is simplified by identifying sorts and constructors of $\T$ with their image in $\pow(\T)$.

\subsection*{Behavior subsystems}
\label{ssec:subsys}



Let $\msf{Th.Gph}\to \T$ be a $\lambda$-theory with internal graph $\mtt{G}=\langle s,t\rangle:\E\to \V,\V$. Then $\sfy\mtt{G}:\pow(\T)$ is the (dependent) type of rewrites over terms. The fiber over each pair is the set of rewrites between terms.
\[\tts,a'\V,b'\V \vdash \mtt{G}(a,b):\Type \quad \quad \quad \quad
\mtt{G}(a,b) = \{\tts\vdash e:a \leadsto b\}\]
This object is the space of all computations in language $\T$. The native type system can be used to construct predicates which specify subgraphs of computations.
\begin{example}
    Let $\msf{Th.Gph}\to \msf{Th}.\uprho\pi$ be the structured $\lambda$-theory of the $\uprho\pi$-calculus (ex.\ \ref{ssec:rho}). In $\msc{L}\pow(\msf{Th}.\uprho\pi)$, suppose we have a name predicate $\alpha:\N\to \Prop$, a process predicate $\varphi:\PP\to \Prop$, and a constructor of predicates $F:[\N\to \Prop]\to [\PP\to \Prop]$.
    Then $\mtt{comm}(\alpha,\varphi,\chi.F):[\mtt{E},\Prop]$ determines communications
    \begin{itemize} 
    \item on channels in namespace $\alpha$
    \item sending data in codespace $\varphi$
    \item and continuing in contexts $\lambda x.c:[\N,\PP]$ such that $\chi(@p)\ra F(\chi)(c[@p/x])$.
    \end{itemize}
    
    This provides highly expressive specification and conditioning of communication on a network. In particular, these predicates could be used to analyze, distribute, or enforce rules for execution on a public computing platform, such as a blockchain.
\end{example}

\subsection*{Modalities}
\label{ssec:modal}
We can express \textit{temporal modalities} to reason about past and future behavior. Applying the ``step'' operators of ex.\ \ref{ex:step}
to a predicate $\varphi:\mtt{V}\to \Prop$ on terms, $\msf{B}_!(\varphi)$ are terms which \textit{possibly} rewrite to $\varphi$, and $\msf{B}_*(\varphi)$ are terms which \textit{necessarily} rewrite to $\varphi$. By iterating, we can form each kind of modality.
\[\begin{array}{ll@{\quad}lcll@{\quad}l}
    \msf{B}_!^\circ(\varphi) := & \exists n'\mbb{N}.\msf{B}_!^n(\varphi) & \text{can become } \varphi && \msf{B}_!^\bullet(\varphi) := & \forall n'\mbb{N}.\msf{B}_!^n(\varphi) & \text{always can become } \varphi\\
    \msf{B}_*^\circ(\varphi) := & \exists n'\mbb{N}.\msf{B}_*^n(\varphi) & \text{will become } \varphi && \msf{B}_*^\bullet(\varphi) := & \forall n'\mbb{N}.\msf{B}_*^n(\varphi) & \text{always } \varphi
\end{array}\]
Similarly for $\msf{F}$, we can condition past behavior. These modalities can also be restricted to subsystems.

\begin{example}
    We can use modalities to express system requirements, such as the capacity to receive and process input on certain channels, or the guarantee to only communicate on certain channels.
    \[\begin{array}{llcll}
        \msf{live}(\alpha) := & \msf{B}_*^\bullet(\tti(\alpha,[\N\to \PP]) \;|\; \PP) & \quad \quad \quad &
        \msf{safe}(\alpha) := & \msf{B}_*^\bullet(\neg[\tti(\neg[\alpha],[\N\to \PP]) \;|\; \PP])
    \end{array}\]
By proving $\tti(n,\lambda x.c):\tti(\N,\chi.\msf{safe})$, we know the program will be secure on the channel it receives.
\end{example}

\subsection*{Behavioral equivalence}
\label{ssec:bisim}

Our rewrite graphs are deterministic, because each edge specifies all data in the term vertices. In operational semantics, rewrites are ``silent reductions'' which occur in a closed system, while \textit{transitions} allow for interaction with the environment. This can be expressed using substitution as pattern-matching, to construct a nondeterministic labelled transition system in which to derive behavioral equivalence.
\begin{example}
\label{ex:obs}
    Processes in the $\uprho\pi$-calculus interact in parallel $-|-$. The basic actions are input and output. To construct the transition system of these observable behaviors, we define interaction contexts.
    \[\msf{obs}:= [\lambda x.x]\lor [ \lambda x.(\tti(\N,\N\to \PP)\;|\; x)]\lor [ \lambda x.(\tto(\N,\PP) \;|\; x)]:[\PP\to \PP]\to \Prop\]
    We can then define the labelled transition system $\msf{act}:\PP,[\PP\to \PP],\PP\to \Prop$ as
    \[p'\PP,\lambda x.c'[\PP\to \PP],q'\PP\vdash \msf{act}(p,\lambda x.c,q):=\mtt{G}(ev[p,\msf{obs}(\lambda x.c)],q).\]
    Hence the predicate which is usually written as
    $p\xr{\lambda x.c}{} q$, meaning ``substituting $p$ into $c$ rewrites to $q$'', we define to be $\exists e:\mtt{G}.\; e:c[p/x]\leadsto q$.
    We can now construct new modalities relative to this observational graph, denoted with $(-)_\msf{act}$.
    
    From this relation, many kinds of behavioral equivalence can be written explicitly as types. For example, bisimulation is the inductive type $\msf{Bisim}:=\mu \varphi.\msf{S}(\varphi)$ for
    \[\begin{array}{lcl}
        \msf{S}(\varphi)(p,q) & := &
        \forall y'\PP.\; \forall \lambda x.c'[\PP,\PP].\; \msf{act}(p,\lambda x.c,y)
        \ra \exists z'\PP.\; \msf{act}(q,\lambda x.c,z) \land \varphi(y,z) \; \land \\
        && \forall z'\PP.\; \forall \lambda x.c'[\PP,\PP].\; \msf{act}(q,\lambda x.c,z)
        \ra \exists y'\PP.\; \msf{act}(p,\lambda x.c,y) \land \varphi(y,z)
    \end{array}\]
    By constructing bisimilarity as a native type, we can reason up to behavioral equivalence.
\end{example}

\subsection*{Refined binding}
\label{ssec:hom}


Hom types provide \textit{refined binding}: using predicates to condition what can be substituted into a context. To do this, we restrict rewrite rules to require that a term satisfies the predicate which the context binds.

\begin{example}
\label{ex:refine}
    In the $\uprho\pi$-calculus, an input process $\tti(n,\lambda x.c)$ receives whatever is sent on the name $n$. We can refine input to receive only data which satisfies a predicate.

    Consider the predicate theory (def.\ \ref{def:predthy}) of the $\uprho\pi$-calculus. For each namespace $\alpha$, define
    \[\mtt{comm}_\alpha:\N,\alpha[@],[\alpha\to \PP]\to \E \quad \quad \mtt{comm}_\alpha(n,p,\lambda x.c): \tto_\alpha(n,p)\vert \tti_\alpha(n,\lambda x.c)\leadsto c[@p/x]\]
    where $\alpha[@]$ is the preimage of $\alpha$ under $@:\PP\to \N$. This extends to polyadic communication.
    
    The \define{refinement} of the $\uprho\pi$-calculus is defined to be the subtheory $\uprho\pi_\omega\subset \omega\Th.\uprho\pi$ in which the only rewrite constructors are $\mtt{comm}_\alpha$ for each namespace. In this theory, $\tti_\alpha:\N,[\alpha\to \PP]\to \PP$ constructs processes which only receive data on $\alpha$.
    
    The namespace $\alpha:\N\to \Prop$ could be a predicate on structured data, a set of trusted addresses, or the implementations of an algorithm. Then $\tti(n,\lambda x'\alpha.p)$ can be understood as a \textit{query} for $\alpha$. In the refined language, we can search by both structure and behavior.
    
\end{example}

\subsection*{Reasoning about contexts}
\label{ssec:ctx}

A ubiquitous question in software is ``what contexts ensure this implication?'' For example, ``where can this protocol be executed without security leaks?'' Hom types provide this expressive power for reasoning contextually in codebases.

By composing the hom type with modalities, we can extend contextual reasoning over term behavior. In particular, $\varphi \rhd \psi := [\varphi,\msf{B}^\circ_*(\psi)]$ are contexts for which substituting $\varphi$ can \textit{eventually} lead to some condition, desired or otherwise.

\begin{example}
    An arrow can be used to detect security leaks: given a trusted channel $a:\N$ and an untrusted $n:\N$, then the following program will not preserve safety on $a$.
    \[\lambda p.(p \;|\; \tto(a,\tti(n,\lambda x.c))): \msf{safe}(a)\rhd \neg[\msf{safe}](a)\]
    
    We can similarly detect if a program may not remain single-threaded. Let $\msf{s.thr}:= \neg [\ttz]\land \neg[\neg[\ttz] \;|\; \neg[\ttz]]$. Then
    \[\lambda p. \tto(a,(p \;|\; q)) : \msf{s.thr}\rhd_\msf{act} \neg[\msf{s.thr}]\]
    where $\rhd_\msf{act}$ is the arrow for the $\msf{act}$ transition system (ex.\ \ref{ex:obs}).
    
    In this way, the process of finding bugs can be automated as a form of type-checking. The query time depends only on the system complexity and the efficiency of the type checker.
\end{example}


    


These kind of predicates in the $\uprho\pi$-calculus have been studied for object capabilities \cite{ocaps}, advocating for better security by determining authority purely through object references.

\begin{example}
    A key concept in concurrency is that of a \textit{race condition}, in which multiple communications on one channel are possible simultaneously.
    \[\begin{array}{ll}
        \msf{race.out} & := \exists n:\N.\; \tto(n,\PP)\;|\;\tto(n,\PP)\;|\;\tti(n,\N.\PP)\;|\;\PP
    \end{array}\]
    We can use native types to design algorithms which identify and manage these conditions. Let $\mtt{comm}(n):= \lambda p. (p \;\vert\; s(\mtt{comm}(n,\PP,\N\to \PP)))$, contexts with potential communication on $n$.
    \[\begin{array}{lclclcl}
        \mtt{comm}(n) & \subseteq & [\tto(n,\PP),\msf{race.out}] & \quad \quad &
        (\mtt{comm}(n) \;|\; \tto(n,\PP)) & \subseteq & \msf{race.out}
    \end{array}\]
    This is useful especially in applications such as blockchain.
\end{example}

\subsection*{Translating across language paradigms}
\label{ssec:translate}


The construction of native type systems is functorial, allowing us to reason across translations. We sketch a simple example of the benefits of relating across programming paradigms.



We give a translation $\tau:\msf{Th.N}\lambda\to \msf{Th}.\pi$ from the name-passing $\lambda$-calculus into the $\pi$-calculus.



\begin{example} Name-passing $\lambda$-calculus  \cite{boudol} (abridged)
\[\begin{array}{rlcrlcrl}
    \V & \text{variables} &&
    \ttt & \text{terms} &&
    \E & \text{rewrites of terms} \; ( + \msf{Th.Cat})\\
\end{array}\]
\[\begin{array}{rlcrlcrl}
    \lam: & [\V\to \ttt]\to \ttt && 
    \var: & \V\to \ttt && 
    \ttC: & \V,\ttt,\ttt\to \ttt\\
    \app: & \ttt,\V\to \ttt &&
    \deff: & \ttt,[\V\to \ttt]\to \ttt\\
\end{array}\]
\[\begin{array} {rlcrl}
    \beta: & [\V\to \ttt],\V\to \E &&
    \beta(Q, y): & \app(\lam(Q),y) \leadsto Q(y)\\
    \phi: & \V,\ttt,\ttt\to \E &&
    \phi(x,Q): & \ttC(x,Q,\var(x))\leadsto Q
\end{array}\]

The name-passing $\lambda$-calculus uses references to avoid copying large data structures. It is a restriction of the $\lambda$-calculus in that terms may only be applied to variables, while it is an enrichment in that it introduces an environment $\deff$ that records binding. There is also a carrier $\ttC$, which serves to transport the recorded binding from its declaration to its use.

The usual $\beta$ reduction splits into two reductions.  The first, denoted $\beta$, replaces variables in a term with other variables.  The second, denoted $\phi$ (for ``fetch''), replaces a variable in head position with the term to which it is bound in the environment.


\end{example}

\begin{example} Polyadic asynchronous $\pi$-calculus  \cite{pical} (abridged)
\[\begin{array}{rlcrlcrl}
    \N & \text{names} && \PP & \text{processes} && \E & \text{rewrites between processes} \; (+ \msf{Th.Cat})
\end{array}\]
\[\begin{array}{rlcrlc}
    \ttz: & 1\to \PP &&
    \mtt{in}_k: & \N,[\N^k\to \PP]\to \PP\\
    -\vert-: & \PP,\PP \to \PP &&
    \mtt{out}_k: & \N, \N^k\to \PP\\
    !: & \PP\to \PP &&
    \nu: & [\N\to \PP]\to \PP & \mbox{syntactic sugar: } \nu x.p \mbox{ means } \nu(\lambda x.p)
\end{array}\]
\[\begin{array}{rlcrl}
    \mtt{comm}_k : & \N, \N^k, [\N^k \to \PP]\to \E &&
    \mtt{comm}_k(n,\vec{a_i},\lambda \vec{y_i}.Q): & \mtt{out}_k(n;\vec{a_i})\vert \mtt{in}_k(n, \lambda \vec{y_i}.Q) \leadsto Q[a_i/y_i]\\
    \mtt{par}_l: & \E, \PP \to \E && \mtt{par}_l(\langle p,e\rangle,q): & p\vert q \leadsto t(e)\vert q \\
    \nu_\tte: & [\N\to \E]\to \E && \nu_\tte x.\rho: & \nu x.s(\rho) \leadsto \nu x.t(\rho)
\end{array}\]

The $\pi$-calculus \cite{polyadic} models concurrent processes which compute via \textit{communication}, or the exchange of ``names''. It is like the $\uprho\pi$-calculus of this paper, without reflection and with two added constructors.
The replication operator $!$ makes infinitely many copies of a process.  The $\nu$ operator introduces a new scope in which a fresh name has been made available to the contained process.
Scopes can expand via scope extrusion to absorb other processes running in parallel with the scope.  


\end{example}

\begin{proposition}
    There is a translation $\interp{-}:\Th.\msf{N}\lambda\to \Th.\pi$ as follows. 
    
    On sorts, $\interp{\V} = \N$, $\interp{\ttt} = [\N\to \PP]$, and $\interp{\mtt{Hom}_\V} = \mtt{Hom}_\PP$.
\end{proposition}
\[\begin{array}{ll}
    \interp{\var}: \N\to [\N\to \PP]\\
    \interp{\var(x)} = \lambda u.\tto_1(x, u)\\\\
    
    \interp{\lam}: [\N\to [\N\to \PP]]\to [\N\to \PP]\\
    \interp{\lam(\lambda x.Q)} = \lambda u.\tti_2(u, \lambda x.\interp{Q})\\\\
    
    \interp{\app}: [\N\to \PP],\N\to [\N\to \PP]\\
    \interp{\app(Q, x)} = \lambda u.\nu v.(\interp{Q}(v)\vert \tto_2(v; x,u))\\\\
    
    \interp{\deff}: [\N\to \PP],[\N\to [\N\to \PP]]\to [\N\to \PP]\\
    \interp{\deff(Q, \lambda x.R)} = \lambda u.\nu x.(\interp{R}(u)\vert !\tti_1(x,\interp{Q}))\\\\
    
    \interp{\ttC}: \N,[\N\to \PP],[\N\to \PP]\to [\N\to \PP]\\
    \interp{\ttC(x, Q, R)} = \lambda u.(\interp{R}(u)\vert \tti_1(x, \interp{Q}))
\end{array}\]

The translation preserves equations and rewrites.
This induces a functor $\pow(\msf{Th}.\pi)\to \pow(\msf{Th.N}\lambda)$, which in turn induces a translation of the native type systems.

A $\pi$-calculus predicate $\varphi:\PP\to \Prop$ contains processes which may involve interaction between agents in a network that is highly nondeterministic. By the translation, ir is mapped to a $\lambda$-calculus predicate by precomposition; this has the effect of restricting $\varphi$ to its ``functional'' processes.

Because $\lambda$-terms have no side-effects and execute deterministically, restricting to functional terms can allow for significant optimization in network computing; e.g.\ agents trying to reach consensus about side effects. A compiler could recognize that a $\pi$-term can be implemented functionally and run the consensus protocol on not the details of the execution but only the result.

\vspace{1em}

These few small examples are only a modest selection of the applications of native type theory. Native types are practical because they are basic: they are made by logic from languages we already use. We encourage the reader to explore what native types can do for you.


\section{Conclusion}
Native type theory is a method to generate expressive type systems for a broad class of languages. We believe that integrating native type systems in software can help to provide a shared framework of higher-order reasoning in everyday computing. Most of the tools necessary for implementation already exist.

\bibliographystyle{eptcs}
\bibliography{ntt}

\begin{thebibliography}{10}
\providecommand{\bibitemdeclare}[2]{}
\providecommand{\surnamestart}{}
\providecommand{\surnameend}{}
\providecommand{\urlprefix}{Available at }
\providecommand{\url}[1]{\texttt{#1}}
\providecommand{\href}[2]{\texttt{#2}}
\providecommand{\urlalt}[2]{\href{#1}{#2}}
\providecommand{\doi}[1]{doi:\urlalt{http://dx.doi.org/#1}{#1}}
\providecommand{\eprint}[1]{arXiv:\urlalt{https://arxiv.org/abs/#1}{#1}}
\providecommand{\bibinfo}[2]{#2}

\bibitemdeclare{misc}{facebook}
\bibitem{facebook}
\emph{\bibinfo{title}{Flow: A Static Type Checker for Javascript}}.
\newblock \urlprefix\url{https://flow.org/}.

\bibitemdeclare{misc}{google}
\bibitem{google}
\emph{\bibinfo{title}{Google Closure Compiler}}.
\newblock \urlprefix\url{https://developers.google.com/closure/compiler}.

\bibitemdeclare{misc}{hoogle}
\bibitem{hoogle}
\emph{\bibinfo{title}{Hoogle}}.
\newblock \urlprefix\url{https://hoogle.haskell.org/}.

\bibitemdeclare{misc}{kframe}
\bibitem{kframe}
\emph{\bibinfo{title}{K Framework}}.
\newblock \urlprefix\url{http://www.kframework.org/}.

\bibitemdeclare{misc}{js-k}
\bibitem{js-k}
\emph{\bibinfo{title}{KJS: A Complete Formal Semantics of JavaScript}}.
\newblock \urlprefix\url{https://github.com/kframework/javascript-semantics}.

\bibitemdeclare{misc}{microsoft}
\bibitem{microsoft}
\emph{\bibinfo{title}{Microsoft TypeScript}}.
\newblock \urlprefix\url{https://www.typescriptlang.org/}.

\bibitemdeclare{misc}{rchain}
\bibitem{rchain}
\emph{\bibinfo{title}{RChain}}.
\newblock \urlprefix\url{https://www.rchain.coop/}.

\bibitemdeclare{book}{awodey-CT}
\bibitem{awodey-CT}
\bibinfo{author}{Steve \surnamestart Awodey\surnameend} (\bibinfo{year}{2010}):
  \emph{\bibinfo{title}{Category Theory}}, \bibinfo{edition}{2nd} edition.
\newblock \bibinfo{publisher}{Oxford University Press, Inc.},
  \bibinfo{address}{USA}.

\bibitemdeclare{article}{naturalModels}
\bibitem{naturalModels}
\bibinfo{author}{Steve \surnamestart Awodey\surnameend} (\bibinfo{year}{2016}):
  \emph{\bibinfo{title}{Natural models of homotopy type theory}}.
\newblock {\sl \bibinfo{journal}{Mathematical Structures in Computer Science}}
  \bibinfo{volume}{28}(\bibinfo{number}{2}), pp. \bibinfo{pages}{241--286},
  \doi{10.1017/s0960129516000268}.

\bibitemdeclare{book}{barendregt}
\bibitem{barendregt}
\bibinfo{author}{H.~P. \surnamestart Barendregt\surnameend}
  (\bibinfo{year}{1984}): \emph{\bibinfo{title}{The Lambda Calculus: Its Syntax
  and Semantics}}.
\newblock \bibinfo{publisher}{Elsevier}.

\bibitemdeclare{inproceedings}{boudol}
\bibitem{boudol}
\bibinfo{author}{G{\'{e}}rard \surnamestart Boudol\surnameend}
  (\bibinfo{year}{1997}): \emph{\bibinfo{title}{The $\uppi$-calculus in direct
  style}}.
\newblock In: {\sl \bibinfo{booktitle}{Proceedings of the 24th {ACM}
  {SIGPLAN}-{SIGACT} symposium on Principles of programming languages -
  {POPL}}}, \bibinfo{publisher}{{ACM} Press}, \doi{10.1145/263699.263726}.

\bibitemdeclare{article}{gats}
\bibitem{gats}
\bibinfo{author}{John \surnamestart Cartmell\surnameend}
  (\bibinfo{year}{1986}): \emph{\bibinfo{title}{Generalised algebraic theories
  and contextual categories}}.
\newblock {\sl \bibinfo{journal}{Annals of Pure and Applied Logic}}
  \bibinfo{volume}{32}, pp. \bibinfo{pages}{209--243},
  \doi{10.1016/0168-0072(86)90053-9}.
\newblock
  \urlprefix\url{https://www.sciencedirect.com/science/article/pii/0168007286900539}.

\bibitemdeclare{article}{coc}
\bibitem{coc}
\bibinfo{author}{Thierry \surnamestart Coquand\surnameend} \&
  \bibinfo{author}{G{\'{e}}rard \surnamestart Huet\surnameend}
  (\bibinfo{year}{1988}): \emph{\bibinfo{title}{The calculus of
  constructions}}.
\newblock {\sl \bibinfo{journal}{Information and Computation}}
  \bibinfo{volume}{76}(\bibinfo{number}{2-3}), pp. \bibinfo{pages}{95--120},
  \doi{10.1016/0890-5401(88)90005-3}.

\bibitemdeclare{book}{crole}
\bibitem{crole}
\bibinfo{author}{Roy~L. \surnamestart Crole\surnameend} (\bibinfo{year}{1994}):
  \emph{\bibinfo{title}{Categories for Types}}.
\newblock \bibinfo{publisher}{Cambridge University Press},
  \doi{10.1017/CBO9781139172707}.

\bibitemdeclare{book}{harper}
\bibitem{harper}
\bibinfo{author}{Robert \surnamestart Harper\surnameend}
  (\bibinfo{year}{2016}): \emph{\bibinfo{title}{Practical Foundations for
  Programming Languages}}, \bibinfo{edition}{2} edition.
\newblock \bibinfo{publisher}{Cambridge University Press},
  \doi{10.1017/CBO9781316576892}.

\bibitemdeclare{misc}{hirsch}
\bibitem{hirsch}
\bibinfo{author}{André \surnamestart Hirschowitz\surnameend},
  \bibinfo{author}{Tom \surnamestart Hirschowitz\surnameend} \&
  \bibinfo{author}{Ambroise \surnamestart Lafont\surnameend}
  (\bibinfo{year}{2020}): \emph{\bibinfo{title}{Modules over monads and
  operational semantics}}.
\newblock \eprint{2012.06530}.

\bibitemdeclare{book}{jacobs}
\bibitem{jacobs}
\bibinfo{author}{B.~\surnamestart Jacobs\surnameend} (\bibinfo{year}{1998}):
  \emph{\bibinfo{title}{Categorical Logic and Type Theory}}.
\newblock \bibinfo{publisher}{Elsevier}, \bibinfo{address}{Amsterdam},
  \doi{10.1016/s0049-237x(98)x8028-6}.

\bibitemdeclare{book}{elephant}
\bibitem{elephant}
\bibinfo{author}{Peter~T. \surnamestart Johnstone\surnameend}
  (\bibinfo{year}{2002}): \emph{\bibinfo{title}{Sketches of an Elephant: A
  Topos Theory Compendium: 2 Volume Set}}.
\newblock \bibinfo{publisher}{Oxford University Press UK}.

\bibitemdeclare{book}{lambek}
\bibitem{lambek}
\bibinfo{author}{J.~\surnamestart Lambek\surnameend} \& \bibinfo{author}{P.~J.
  \surnamestart Scott\surnameend} (\bibinfo{year}{1986}):
  \emph{\bibinfo{title}{Introduction to Higher Order Categorical Logic}}.
\newblock \bibinfo{publisher}{Cambridge University Press},
  \bibinfo{address}{USA}.

\bibitemdeclare{article}{hyper}
\bibitem{hyper}
\bibinfo{author}{F.~William \surnamestart Lawvere\surnameend}
  (\bibinfo{year}{1969}): \emph{\bibinfo{title}{Adjointness in Foundations}}.
\newblock {\sl \bibinfo{journal}{dialectica}}
  \bibinfo{volume}{23}(\bibinfo{number}{3-4}), pp. \bibinfo{pages}{281--296},
  \doi{10.1111/j.1746-8361.1969.tb01194.x}.

\bibitemdeclare{book}{working}
\bibitem{working}
\bibinfo{author}{Saunders \surnamestart MacLane\surnameend}
  (\bibinfo{year}{1971}): \emph{\bibinfo{title}{Categories for the Working
  Mathematician}}.
\newblock \bibinfo{publisher}{Springer New York},
  \doi{10.1007/978-1-4612-9839-7}.

\bibitemdeclare{book}{sheavesinGL}
\bibitem{sheavesinGL}
\bibinfo{author}{Saunders \surnamestart MacLane\surnameend} \&
  \bibinfo{author}{Ieke \surnamestart Moerdijk\surnameend}
  (\bibinfo{year}{1994}): \emph{\bibinfo{title}{Sheaves in Geometry and
  Logic}}.
\newblock \bibinfo{publisher}{Springer New York},
  \doi{10.1007/978-1-4612-0927-0}.

\bibitemdeclare{incollection}{martin-lof}
\bibitem{martin-lof}
\bibinfo{author}{Per \surnamestart Martin-Löf\surnameend}
  (\bibinfo{year}{1998}): \emph{\bibinfo{title}{An intuitionistic theory of
  types}}.
\newblock In: {\sl \bibinfo{booktitle}{Twenty Five Years of Constructive Type
  Theory}}, \bibinfo{publisher}{Oxford University Press},
  \doi{10.1093/oso/9780198501275.003.0010}.

\bibitemdeclare{article}{refine}
\bibitem{refine}
\bibinfo{author}{Paul-Andr{\'{e}} \surnamestart Melli{\`{e}}s\surnameend} \&
  \bibinfo{author}{Noam \surnamestart Zeilberger\surnameend}
  (\bibinfo{year}{2015}): \emph{\bibinfo{title}{Functors are Type Refinement
  Systems}}.
\newblock {\sl \bibinfo{journal}{{ACM} {SIGPLAN} Notices}}
  \bibinfo{volume}{50}(\bibinfo{number}{1}), pp. \bibinfo{pages}{3--16},
  \doi{10.1145/2775051.2676970}.

\bibitemdeclare{incollection}{namespace}
\bibitem{namespace}
\bibinfo{author}{L.~G. \surnamestart Meredith\surnameend} \&
  \bibinfo{author}{Matthias \surnamestart Radestock\surnameend}
  (\bibinfo{year}{2005}): \emph{\bibinfo{title}{Namespace Logic: A Logic for a
  Reflective Higher-Order Calculus}}.
\newblock In: {\sl \bibinfo{booktitle}{Trustworthy Global Computing}},
  \bibinfo{publisher}{Springer Berlin Heidelberg}, pp.
  \bibinfo{pages}{353--369}, \doi{10.1007/11580850_19}.

\bibitemdeclare{article}{rhocal}
\bibitem{rhocal}
\bibinfo{author}{L.G. \surnamestart Meredith\surnameend} \&
  \bibinfo{author}{Matthias \surnamestart Radestock\surnameend}
  (\bibinfo{year}{2005}): \emph{\bibinfo{title}{A Reflective Higher-order
  Calculus}}.
\newblock {\sl \bibinfo{journal}{Electronic Notes in Theoretical Computer
  Science}} \bibinfo{volume}{141}(\bibinfo{number}{5}), pp.
  \bibinfo{pages}{49--67}, \doi{10.1016/j.entcs.2005.05.016}.

\bibitemdeclare{misc}{ocaps}
\bibitem{ocaps}
\bibinfo{author}{Lucius~G \surnamestart Meredith\surnameend},
  \bibinfo{author}{Mike \surnamestart Stay\surnameend} \&
  \bibinfo{author}{Sophia \surnamestart Drossopoulou\surnameend}
  (\bibinfo{year}{2013}): \emph{\bibinfo{title}{Policy as Types}}.
\newblock \eprint{1307.7766}.

\bibitemdeclare{incollection}{polyadic}
\bibitem{polyadic}
\bibinfo{author}{Robin \surnamestart Milner\surnameend} (\bibinfo{year}{1993}):
  \emph{\bibinfo{title}{The Polyadic $\uppi$-Calculus: a Tutorial}}.
\newblock In: {\sl \bibinfo{booktitle}{Logic and Algebra of Specification}},
  \bibinfo{publisher}{Springer Berlin Heidelberg}, pp.
  \bibinfo{pages}{203--246}, \doi{10.1007/978-3-642-58041-3_6}.

\bibitemdeclare{article}{wtypes}
\bibitem{wtypes}
\bibinfo{author}{Ieke \surnamestart Moerdijk\surnameend} \&
  \bibinfo{author}{Erik \surnamestart Palmgren\surnameend}
  (\bibinfo{year}{2000}): \emph{\bibinfo{title}{Wellfounded trees in
  categories}}.
\newblock {\sl \bibinfo{journal}{Annals of Pure and Applied Logic}}
  \bibinfo{volume}{104}(\bibinfo{number}{1-3}), pp. \bibinfo{pages}{189--218},
  \doi{10.1016/s0168-0072(00)00012-9}.

\bibitemdeclare{article}{pical}
\bibitem{pical}
\bibinfo{author}{Davide \surnamestart Sangiorgi\surnameend}
  (\bibinfo{year}{2000}): \emph{\bibinfo{title}{Communicating and Mobile
  Systems: the $\uppi$-calculus,}}.
\newblock {\sl \bibinfo{journal}{Science of Computer Programming}}
  \bibinfo{volume}{38}(\bibinfo{number}{1-3}), pp. \bibinfo{pages}{151--153},
  \doi{10.1016/s0167-6423(00)00008-3}.

\bibitemdeclare{misc}{scott}
\bibitem{scott}
\bibinfo{author}{Dana~S. \surnamestart Scott\surnameend}
  (\bibinfo{year}{1980}): \emph{\bibinfo{title}{Relating Theories of the Lambda
  Calculus}}.
\newblock
  \urlprefix\url{https://www.andrew.cmu.edu/user/awodey/dump/Scott/ScottRelating.pdf}.

\bibitemdeclare{article}{coqTypeCheck}
\bibitem{coqTypeCheck}
\bibinfo{author}{Matthieu \surnamestart Sozeau\surnameend},
  \bibinfo{author}{Simon \surnamestart Boulier\surnameend},
  \bibinfo{author}{Yannick \surnamestart Forster\surnameend},
  \bibinfo{author}{Nicolas \surnamestart Tabareau\surnameend} \&
  \bibinfo{author}{Th\'{e}o \surnamestart Winterhalter\surnameend}
  (\bibinfo{year}{2019}): \emph{\bibinfo{title}{Coq Coq Correct! Verification
  of Type Checking and Erasure for Coq, in Coq}}.
\newblock {\sl \bibinfo{journal}{Proc. ACM Program. Lang.}}
  \bibinfo{volume}{4}(\bibinfo{number}{POPL}), \doi{10.1145/3371076}.

\bibitemdeclare{inproceedings}{cosmos}
\bibitem{cosmos}
\bibinfo{author}{Ross \surnamestart Street\surnameend} (\bibinfo{year}{1974}):
  \emph{\bibinfo{title}{Elementary cosmoi I}}.
\newblock In \bibinfo{editor}{Gregory~M. \surnamestart Kelly\surnameend},
  editor: {\sl \bibinfo{booktitle}{Category Seminar}},
  \bibinfo{publisher}{Springer Berlin Heidelberg}, \bibinfo{address}{Berlin,
  Heidelberg}, pp. \bibinfo{pages}{134--180},
  \doi{10.1016/0022-4049(72)90019-9}.

\bibitemdeclare{article}{burstall}
\bibitem{burstall}
\bibinfo{author}{Andrzej \surnamestart Tarlecki\surnameend},
  \bibinfo{author}{Rod~M. \surnamestart Burstall\surnameend} \&
  \bibinfo{author}{Joseph~A. \surnamestart Goguen\surnameend}
  (\bibinfo{year}{1991}): \emph{\bibinfo{title}{Some fundamental algebraic
  tools for the semantics of computation: Part 3. indexed categories}}.
\newblock {\sl \bibinfo{journal}{Theoretical Computer Science}}
  \bibinfo{volume}{91}(\bibinfo{number}{2}), pp. \bibinfo{pages}{239 -- 264},
  \doi{10.1016/0304-3975(91)90085-G}.
\newblock
  \urlprefix\url{http://www.sciencedirect.com/science/article/pii/030439759190085G}.

\bibitemdeclare{inproceedings}{opsem}
\bibitem{opsem}
\bibinfo{author}{D.~\surnamestart Turi\surnameend} \&
  \bibinfo{author}{G.~\surnamestart Plotkin\surnameend}:
  \emph{\bibinfo{title}{Towards a mathematical operational semantics}}.
\newblock In: {\sl \bibinfo{booktitle}{Proceedings of Twelfth Annual {IEEE}
  Symposium on Logic in Computer Science}}, \bibinfo{publisher}{{IEEE} Comput.
  Soc}, \doi{10.1109/lics.1997.614955}.

\end{thebibliography}

\end{document}